\documentclass[preprint,showpacs]{revtex4-1}
\usepackage{amsmath}
\usepackage{amsfonts}
\usepackage{graphicx}
\usepackage{amsmath,amssymb,graphicx,bm}
\usepackage{epsfig,subfigure}
\usepackage{epsfig}
\begin{document} 
\title{Diffusion coefficients preserving long-time correlations: Consequences on  the Einstein relation and on entanglement  in a  bosonic Bogoliubov system}
\author{Yamen Hamdouni}
\email{hamdouniyamen@gmail.com; hamdouni.yamen@umc.edu.dz}
\affiliation{Department of Physics,  Mentouri University, Constantine, Algeria}
\begin{abstract}
We analytically derive the diffusion coefficients that drive a system of $N$ coupled harmonic oscillators to an equilibrium state exhibiting persistent correlations. It is shown that the main effect of the latter consists in a renormalization of the natural frequencies and the friction  coefficients of the oscillators. We find  that the Einstein relation may be satisfied at low temperatures with frequency-dependent effective friction coefficients, provided that the physical constraints are fulfilled.  We also investigate the entanglement evolution in a bipartite bosonic Bogoliubov system initially prepared in a thermal squeezed state. It is found that, in contrast to what  one may expect, strong coupling slows down the entanglement sudden death, and for initially separable states, entanglement generation may occur.  
\end{abstract}

\maketitle
\section{Introduction}

Quantum systems out of equilibrium exhibit rich features and are the origin of several intriguing results and concepts in quantum and statistical physics \cite{zwanz}. The significance and complexity of such systems solicited a great deal of interest from different perspectives and backgrounds. In particular, the relaxation of a system to its steady state due to dissipation is a fundamental problem that stimulated the advancement of the theoretical investigation of nonequilibrium phenomena.  Historically,  the classical study of  problems related to dissipation gave rise to two different but equivalent approaches, namely the Langevin and the Fokker-Planck equations \cite{fock}. In this regard, the process of friction, which is often introduced  phenomenologically, as responsible of the exchange of energy between the system and its surrounding,  plays a central role.  

Generally speaking, friction is incorporated in the equations of motion by introducing the so-called memory-friction kernel. The latter describes the dependence of the dynamics at a given instant on the properties of the relevant variables at earlier times, and is solely related to the stochastic nature of the heat reservoir with which the system is interacting \cite{weiss}. The Markov approximation neglects the memory effects \cite{gardiner}, and as a result the kernel reduces to a constant friction coefficient. This approximation holds well when the coupling of the system to the reservoir is weak. Moreover, dissipation is often associated with transport phenomena, where diffusion plays a central role. The fluctuation-dissipation  theorem relates the friction coefficient  to  an other important parameter, namely, the so-called  diffusion coefficient. For example, the diffusion coefficient, through Fick's law, relates the time change of the density of particles to its spatial variation. 

The extension of the concepts of friction and dissipation to the quantum domain turns out to be nontrivial. In particular, the loss of energy implies that the dynamics is basically nonunitary \cite{semigroup}.   A well established and widely adopted approach to deal with this kind of problems, consists in regarding the system of interest as part of a larger system and to apply the usual quantization procedure to the full system \cite{petr}. Then, the properties of the target subsystem may be obtained by disregarding the other degrees of freedom, which are refereed to the heat bath. This is translated in a mathematical language by taking the partial trace over the reservoir variables. The outcome depends only on the degrees of freedom of the system. The quantity of interest is the reduced density matrix, the derivation of which  does not yield in general analytically solvable  master equations. Some approximations and simplifications are generally employed among which the Markovian approximation, valid for weak system-reservoir coupling, is the most common one. Lindblad \cite{lind1,lind2} came out with the most general form of Markovian  master equations fulfilling the requirements of the physically acceptable evolution of the reduced density matrix, notably the requirement of complete positivity. The Lindblad equation provides an axiomatic approach that has been used in many contexts  in particular for the harmonic oscillator \cite{sandu}-\cite{hamd2}.

In \cite{hamd1}  the multi-dimensional diffusion coefficients for a system of $N$ coupled harmonic oscillators were derived assuming that the system completely thermalizes. The aim of the present work is to investigate the effect of persistent correlations in the steady state on the diffusion coefficients. 
 In section~\ref{sec2} we introduce the model Hamiltonian along with the equilibrium state. We then derive analytically the expressions of the diffusion coefficients leading to the steady state. This is followed by a discussion of the validity of the Einstein relation.  Section~\ref{sec3} deals with the evolution of entanglement in a Bogoliubov bosonic system, where the influence of the coupling constant is investigated. We end the paper with a brief conclusion.

\section{Effect of persistent  steady state correlations on the diffusion coefficients\label{sec2}}

\subsection{System Hamiltonian}
We consider a system of $N$ coupled harmonic oscillators and we denote by   $m_k$ and $\omega_k$ the  mass and the natural frequency of oscillator number $k$.  The positions and momenta operators of  the  oscillators satisfy the canonical commutation relations 
 \begin{equation}\label{can}
[\hat q_k,\hat p_j]=i\hbar\delta_{kj},\qquad [\hat q_k,\hat q_j]=[\hat p_k,\hat p_j]=0.
\end{equation}
The Hamiltonian of the system is given by
\begin{equation}
\hat  H=\hat H_0+\hat H_I \label{orig}
\end{equation}
where $\hat H_0$ describes a set of $N$ independent harmonic oscillators exhibiting position-momentum coupling with strengths $\mu_{kk}$:
\begin{equation}
\hat H_0 =\sum\limits_{k=1}^N \Bigl( \frac{\hat{p}_k^2}{2m_k}+ \frac{1}{2}m_k\omega_k^2 \hat{q}_k^2+\frac{\mu_{kk}}{2}(\hat{ p}_k\hat{q}_k+  \hat{q}_k\hat{p}_k)\Bigl),
\end{equation}
whereas $\hat H_I$ takes into account the coupling between the different subsystems. The latter is assumed  bilinear in the position and momenta operators; explicitly we have
\begin{equation}
\hat H_I=\frac{1}{2}\sum\limits_{k\neq j}^N(\nu_{kj} \hat{q}_k\hat{q}_j+\kappa_{kj}\hat{p}_k\hat{p}_j)+ \sum\limits_{k\neq j}^N \mu_{kj} \hat{p}_k \hat{q}_j,
\end{equation}
 where $\mu_{kj}$, $\nu_{kj}$ and $\kappa_{kj}$ designate  the coupling constants which satisfy the Onsager relations $\nu_{kj}=\nu_{jk}$ and $ \kappa_{kj}=\kappa_{jk}$ ~\cite{onsager}.  Moreover, the classical analog of $\hat H$ is supposed  to be a positive definite quadratic form, which ascertains that the system represents effectively a set of attractive harmonic oscillators.
 
 The aim is to investigate the effect of dissipation on the system of oscillators. To this end we assume that the latter is coupled to a heat reservoir, whose characteristic relaxation time is much smaller than any relevant  time constant associated with the evolution of the system of harmonic oscillators. The above  assumption is mostly verified in the Markovian approximation for which memory effects of the heat bath are irrelevant. Under the above conditions, the evolution of the density matrix of the system verifies 
a Markovian master equation in the Lindblad form, namely,

\begin{equation}
\frac{d\hat\rho(t)}{dt}=-\frac{i}{\hbar}[\hat H,\hat\rho(t)]+\frac{1}{2\hbar}\sum_\ell([\hat V_\ell \hat\rho(t),\hat V_\ell^\dag]+[\hat V_\ell, \hat\rho(t)\hat V_\ell^\dag]),\label{mas1}
\end{equation}
where the operators $\hat V_\ell$ depend solely on the degrees of freedom of the system, but not on those associated with the heat bath. It should be stressed that most of the Markovian master equations found in the literature are special cases of the general form (\ref{mas1}). The dual Lindblad equation for the arbitrary operator $\hat A$ is given by
\begin{equation}
 \frac{d\hat A(t)}{dt}=\frac{i}{\hbar}[\hat H,\hat A(t)]+\frac{1}{2\hbar}\sum_\ell([\hat V_\ell^\dag[ \hat A(t),\hat V_\ell]+[\hat V_\ell^\dag, \hat A(t)]\hat V_\ell]).\label{mas3}
\end{equation}
The general expressions  of the generators  $\hat V_\ell$  read
\begin{equation}
\hat V_\ell=\sum_j (a^\ell_j \hat p_j+b_j^\ell \hat q_j),\quad \hat V^\dag_\ell=\sum_j (a^{\ell*}_j \hat p_j+b_j^{\ell*} \hat q_j),  
\end{equation}
which are linear in the degrees of freedom of the oscillators. Notice that in the above combinations the coefficients   $ a^\ell_j$ and $b_j^\ell$ are  complex numbers.
Let us introduce the new parameters
\begin{eqnarray}
D_{q_kq_j}&=&\frac{\hbar}{2}{\rm Re}\sum_\ell a^{\ell*}_k a^\ell_j, \quad  D_{p_kp_j}=\frac{\hbar}{2}{\rm Re}\sum_\ell b^{\ell*}_k b^\ell_j, \label{coef1}\\
D_{q_kp_j}&=&-\frac{\hbar}{2}{\rm Re}\sum_\ell a^{\ell*}_k b^\ell_j, \quad \lambda_{kj}=-{\rm Im}\sum_\ell a^{\ell*}_k b^\ell_j,\label{coef2}\\
\alpha_{kj}&=&-{\rm Im}\sum_\ell a^{\ell*}_k a^\ell_j,\quad \eta_{kj}=-{\rm Im}\sum_\ell b^{\ell*}_k b^\ell_j.\label{coef3}
\end{eqnarray}
Physically speaking,  $D_{q_kq_j}$, $D_{p_kp_j}$ and $D_{q_kp_j}$  denote the diffusion coefficients, whereas $\lambda_{kj}$ designate the friction coefficients. Moreover, the covariance of any couple of the system operators $\hat A$ and $\hat B$ is defined by
\begin{equation}
\sigma_{ F G}(t)=\frac{1}{2}{\rm tr}\Bigl(\hat \rho\{\hat F(t),\hat G(t)\}\Bigl)-{\rm tr}\Bigl(\hat \rho\  \hat F(t)\Bigl) {\rm tr}\Bigl(\hat \rho\  \hat G(t)\Bigl).
\end{equation}   
 
It can be shown that the covariance matrix of the set of operators $\hat q_k$, $\hat p_k$ evolves in the course of the time according to
\begin{equation}
 \frac{d\sigma(t)}{dt}=M\sigma(t)+\sigma(t)M^T+2D,\label{evol2}
\end{equation}
where
\begin{eqnarray} 
 M= {\footnotesize \begin{pmatrix}
        -\lambda_{11}+\mu_{11}&\tfrac{1}{m1} & -\lambda_{12}+\mu_{12}&-\alpha_{12}+\kappa_{12} &\cdots  &  \cdots &-\lambda_{1N}+\mu_{1N}& -\alpha_{1N}+\kappa_{1N} \\
        - m_1 \omega_1^2 &-\lambda_{11}-\mu_{11}&\eta_{12}-\nu_{12}&-\lambda_{21}-\mu_{21}& \cdots &  \cdots &\eta_{1N}-\nu_{1N}&-\lambda_{N1}-\mu_{N1}\\
        -\lambda_{21}+\mu_{21}&\alpha_{12}+\kappa_{12}&-\lambda_{22}+\mu_{22}&\tfrac{1}{m_2} & \cdots & \cdots & -\lambda_{2N}+\mu_{2N} &-\alpha_{2N}+\kappa_{2N} \\
       -\eta_{12}-\nu_{12}&-\lambda_{12}-\mu_{12}& - m_2 \omega_2^2 &-\lambda_{22}-\mu_{22}& \cdots &  \cdots &\eta_{2N}-\nu_{2N} &-\lambda_{N2}-\mu_{N2}\\ 
       \vdots & \vdots & \vdots & \vdots &\vdots &\vdots &\vdots & \vdots\\
       \vdots & \vdots & \vdots & \vdots & \vdots &   \vdots & \vdots & \vdots\\
       -\lambda_{N1}+\mu_{N1} &\alpha_{1N}+\kappa_{1N}&-\lambda_{N2}+\mu_{N2} &  \cdots & \cdots & \cdots &-\lambda_{NN}+\mu_{NN}& \tfrac{1}{m_N}\\
       -\eta_{1N}-\nu_{1N}& -\lambda_{1N}-\mu_{1N}&-\eta_{2N}-\nu_{2N}&  \cdots & \cdots & \cdots &-m_{N}\omega_N^2 & -\lambda_{NN}-\mu_{NN}
        \label{matr}
        \end{pmatrix}},
\end{eqnarray}
 and $D$ is the diffusion matrix:
 \begin{equation}
D=\begin{pmatrix}D_{q_1q_1}&D_{q_1p_1}&D_{q_1q_2}&D_{q_1p_2}&\cdots \cdots &D_{q_1q_N}&D_{q_1p_N}\\
D_{p_1q_1}&D_{p_1p_1}&D_{p_1q_2}&D_{p_1p_2}&\cdots \cdots&D_{p_1q_N}&D_{p_1p_N}\\
\vdots&\vdots&  \vdots&\vdots &\cdots \cdots& \vdots&\vdots\\\vdots&\vdots&  \vdots&\vdots &\cdots \cdots& \vdots&\vdots\\D_{q_Nq_1}&D_{q_Np_1}&D_{q_N q_2}&D_{q_Np_2}&\cdots \cdots &D_{q_Nq_N}&D_{q_Np_N}\\
D_{p_Nq_1}&D_{p_Np_1}&D_{p_N q_2}&D_{p_Np_2}&\cdots \cdots&D_{p_Nq_N}&D_{p_Np_N}\\
\end{pmatrix}.
\end{equation}
The solution for the covariance matrix then reads as
\begin{equation}
  \sigma(t)=\exp(M t)(\sigma(0)-\tilde\sigma)\exp(M t)^T+\tilde\sigma,
  \end{equation}
   provided that the following condition is satisfied:
  \begin{equation}
  M\tilde\sigma+\tilde\sigma M^T+2D=0.\label{maeq}
  \end{equation}
  \subsection{Equilibrium state}
We assume that the system evolves eventually to the equilibrium  Gibbs state
\begin{equation}
\hat \rho_{\rm{eq}}=\exp(-\beta \hat H_{\rm{eq}})/Z, \qquad Z=\mathrm{tr} \exp(-\beta \hat H_{\rm{eq}}) \label{gibbs},
 \end{equation}
 where 
 \begin{equation}
 \hat H_{\rm{eq}}=\sum\limits_{k=1}^N \Bigl( \frac{\hat{p}_k^2}{2m_k}+ \frac{1}{2}m_k\omega_k^2 \hat{q}_k^2+\frac{\tilde\mu_{kk}}{2}(\hat{ p}_k\hat{q}_k+  \hat{q}_k\hat{p}_k)\Bigl),
 \end{equation}
 and  $Z$ denotes the partition function. The state $\hat \rho_{\rm{eq}}$ clearly retains the position-momentum correlation of each oscillator. The existence of such state implies that 
\begin{equation}
 \lim_{t\to\infty} \exp(M t)=0. \label{relax}
\end{equation}
 Therefore, all the eigenvalues of the matrix $M$ should possess negative real parts, whence  
 \begin{equation}
\tilde\sigma=\sigma(\infty).
\end{equation}
  The coupling constant $\tilde\mu_{kk}$ entering the expression of the asymptotic state  may differ from $\mu_{kk}$ of the original Hamiltonian, which  will  be the case in the subsequent discussion unless stated otherwise. The classical counterpart of $\hat  H_{\rm{eq}}$ is also assumed to be positive definite.  This amounts to assuming that  $\omega_k > \tilde\mu_{kk}$, which  ensures the existence of a stable equilibrium state. Indeed, from a classical perspective the  function $ H_k(q_k,p_k)=p_k^2/(2m_k)+(1/2) m_k \omega_k^2 q_k^2+\tilde\mu_{kk} q_kp_k$ is represented by a two dimensional surface in the phase space $(q_k,p_k)$.  It has a global minimum at the origin when $\omega_k>\tilde\mu_{kk}$, and the equilibrium in this case is stable. It posses a global maximum at the origin when $\omega_k<\tilde\mu_{kk}$ and the state is inherently instable, which we will exclude from our discussion. 
  
  In choosing the equilibrium state (\ref{gibbs}), we have assumed that condition (\ref{relax}) is satisfied and  the dynamics is thus relaxing, implying that the steady state  exists and  is unique and stable~\cite{nieg}. This corresponds to many interesting physical problems, as is the case  for example in  equilibration processes in heavy-ion collisions (e.g. see~\cite{adam2,hamd1}), where the model parameters are fitted to data,   and the validity of the master equation is verified on the basis that it reproduces the experimental findings. The results reported in this work are thus relevant for a wide range of quadratic systems whose   equilibrium states are  known to be  unique and stable thermal states, at least phenomenologically. 
  
  In general, the study of  the uniqueness and the stability of the asymptotic state of the Lindblad equation is a fundamental problem in the field of quantum control, whose full solution remains open.  Nevertheless, for the system under investigation there exists a number of results that serve as criteria to ensure the uniqueness of the  asymptotic state~\cite{spohn1,spohn2,schrimer,albert}. For our purpose,  we first introduce the  annihilation and  creation operators $\hat a_k$ and $\hat a^\dag_k$   of the harmonic oscillators, and then  rewrite the Lindblad equation (\ref{mas1}) in terms of these new operators. In doing so we obtain terms having the form ${\mathcal D}[\hat a_k]\hat\rho(t)=\hat a_k\hat \rho(t)\hat a^\dag_k-\frac{1}{2}(\hat a^\dag_k  \hat a_k\hat \rho(t)+\hat \rho(t)\hat a^\dag_k  \hat a_k)$. Hence by  condition 3  of reference~\cite{schrimer}, which states that, since the annihilation operator for the harmonic oscillator is similar to a Jordan matrix, there exist no two orthogonal invariant subspaces for the Hilbert space of the system; the asymptotic state is thus unique and stable.  Of great significance are the results obtained by Spohn~\cite{spohn1,spohn2}, that make use of the notions of commutant and bicommutant. A  system similar to the one we are dealing with has been investigated in~\cite{rossi}, where the harmonic oscillators are coupled  to an environment constructed from several  heat baths. By Spohn's theorem, it turns out that the asymptotic state is unique. This seems to be a general feature of the harmonic oscillators coupled to thermal reservoirs~\cite{albert}. 
  
  In the subsequent discussion, we assume that the model parameters, in particular the friction coefficients are such that condition  (\ref{relax}) is ascertained, and the dynamics is relaxing.
Afterwards, inserting the density matrix $\hat\rho_{\rm{eq}}$  into the master equation~(\ref{mas1}) yields
\begin{equation}
 e^{\beta \hat H_{\rm{eq}}} \hat H  e^{-\beta \hat H_{\rm{eq}}}-\hat H=\frac{1}{2i}\sum_\ell\Bigl(2e^{\beta \hat H_{\rm{eq}}} \hat V_\ell e^{-\beta \hat H_{\rm{eq}}} \hat V_\ell^\dag-e^{\beta \hat H_{\rm{eq}}}\hat V_\ell^\dag \hat V_\ell e^{-\beta \hat H_{\rm{eq}}}-\hat V_\ell^\dag \hat V_\ell \Bigr)\label{two}.
\end{equation}
Taking into account the Baker-Campbell-Hausdorff formula
\begin{eqnarray}
e^{\beta \hat H_{\rm{eq}}} \hat A  e^{-\beta \hat H_{\rm{eq}}}=
\hat A-\frac{\beta}{1!}[\hat A,\hat H_{\rm{eq}}]+\frac{\beta^2}{2!}[[\hat A,\hat H_{\rm{eq}}],\hat H_{\rm{eq}}]-\frac{\beta^3}{3!}[[[\hat A,\hat H_{\rm{eq}}],\hat H_{\rm{eq}}],\hat H_{\rm{eq}}]+\cdots\Bigr\}, \label{super}
\end{eqnarray}
we obtain for the operator $\hat q_k$:
\begin{eqnarray}
 e^{\beta \hat H_{\rm{eq}}} \hat q_k e^{-\beta \hat H_{\rm{eq}}}&=&\hat q_k-i\hbar \tilde\mu_{kk} \hat q_k-\frac{i \hbar \beta}{m_k}\hat p_k+\frac{\beta^2}{2}(-\hbar^2{\tilde\mu_{kk}}^2\hat q_k-\frac{\hbar^2  \tilde\mu_{kk}}{m_k} \hat p_k+\hbar^2 \omega_k^2 \hat q_k+\frac{\hbar^2  \tilde\mu_{kk}}{m_k} \hat p_k)\nonumber \\&-&\frac{\hbar^2\beta^3}{6}(\omega_k- \tilde\mu_{kk})(i\hbar  \tilde\mu_{kk} \hat q_k+\frac{i \hbar \beta}{m_k}\hat p_k)+\cdots
\end{eqnarray}
This gives
\begin{eqnarray}
 e^{\beta \hat H_{\rm{eq}}} \hat q_k e^{-\beta \hat H_{\rm{eq}}}&=&\hat q_k\Bigl[\cosh\bigl(\hbar\beta\sqrt{\omega_k^2-{\tilde\mu_{kk}}^2}\bigr)-\frac{i\tilde\mu_{kk}}{\sqrt{\omega_k^2-{\tilde\mu_{kk}}^2}}\sinh\bigl(\hbar\beta\sqrt{\omega_k^2-{\tilde\mu_{kk}}^2}\bigr)\Bigr]\nonumber \\&-&\frac{i}{m_k\sqrt{\omega_k^2-{\tilde\mu_{kk}}^2}} \sinh\bigl(\hbar\beta\sqrt{\omega_k^2-{\tilde\mu_{kk}}^2}\bigr) \hat p_k.\label{tor1}
\end{eqnarray}
Similarly, we obtain for $\hat p_k$

\begin{eqnarray}
 e^{\beta \hat H_{\rm{eq}}} \hat p_k e^{-\beta \hat H_{\rm{eq}}}&=&\hat p_k\Bigl[\cosh\bigl(\hbar\beta\sqrt{\omega_k^2-{\tilde\mu_{kk}}^2}\bigr)+\frac{i\tilde\mu_{kk}}{\sqrt{\omega_k^2-{\tilde\mu_{kk}}^2}}\sinh\bigl(\hbar\beta\sqrt{\omega_k^2-{\tilde\mu_{kk}}^2}\bigr)\Bigr]\nonumber \\&+&\frac{im_k \omega_k^2}{\sqrt{\omega_k^2-{\tilde\mu_{kk}}^2}} \sinh\bigl(\hbar\beta\sqrt{\omega_k^2-{\tilde\mu_{kk}}^2}\bigr) \hat q_k. \label{tor2}
\end{eqnarray}
From a quantum mechanical point of view the expectation values of the operators $\hat q_k$ and $\hat p_k$ in the steady state clearly vanish due to the quadratic nature of $\hat H_{\rm{eq}}$. On the other hand we can write  for example:
 \begin{equation}
  \mathrm{tr} \hat q^2 e^{-\beta \hat H_{\rm{eq}}}=\mathrm {tr} e^{\beta \hat H_{\rm{eq}}}  \hat q e^{-\beta \hat H_{\rm{eq}}}  e^{\beta \hat H_{\rm{eq}}}  \hat q e^{-\beta \hat H_{\rm{eq}}} e^{-\beta \hat H_{\rm{eq}}},
 \end{equation}
which must  be positive and monotonic with respect to the temperature in the steady state. The hyperbolic functions displayed in equations (\ref{tor1}) and (\ref{tor2}) should not display oscillatory behavior, and consequently the quantity under  the square root sign should be positive, that is $\omega_k> \tilde\mu_{kk}$, which corresponds to the stable equilibrium state. 

\subsection{Analytical expressions}

 By inserting equations (\ref{tor1}) and (\ref{tor2}) into equation (\ref{two}), and using the coefficients defined above we obtain:
 
 \begin{eqnarray}
 m_k m_j \omega_k \omega_j  D_{q_kq_j}+  D_{p_kp_j}&=&\frac{1}{2} \omega_k \omega_j\Phi_{kj}+2\Psi_{kj}\label{set1}\\
 m_k \omega_k  D_{q_kp_j}+  m_j \omega_j D_{q_jp_k}&=&\frac{1}{2} \omega_k \omega_j\Phi_{kj}-2\Psi_{kj}\label{set2}\\
 -m_k m_j \omega_k \omega_j  D_{q_kq_j}+  D_{p_kp_j}&=& \omega_k\Gamma_{kj}+ \omega_j\Gamma_{jk}\label{set3}\\
  m_k \omega_k  D_{q_kp_j}-  m_j \omega_j D_{q_jp_k}&=& \omega_k\Gamma_{kj}- \omega_j\Gamma_{jk}, \label{set4}
 \end{eqnarray}
where
\begin{eqnarray}
 \Phi_{kj}&=&\frac{\hbar}{4}\Biggl(\frac{\omega_k-{\tilde\mu_{kk}}}{\omega_k \sqrt{\omega_k^2-{\tilde\mu_{kk}}^2}}\Biggl)\Bigl[ \frac{1}{\omega_j}(\eta_{kj}+\nu_{kj})- m_km_j\omega_k (\alpha_{kj} +\kappa_{kj})\nonumber \\&+&\frac{m_k \omega_k}{ \omega_j} (\lambda_{kj}+\mu_{kj})+m_j(\lambda_{jk}-\mu_{jk})\Biggr]\coth\Biggl(\frac{\hbar\beta}{2} \sqrt{\omega_k^2-{\tilde\mu_{kk}}^2}\Biggr)\nonumber \\
 &+& \frac{\hbar}{4}\Biggl(\frac{\omega_j-{\tilde\mu_{jj}}}{\omega_j \sqrt{\omega_j^2-{\tilde\mu_{jj}}^2}}\Biggl)\Bigl[ \frac{1}{\omega_k}(\eta_{jk}+\nu_{kj})- m_km_j\omega_j (\alpha_{jk} +\kappa_{kj})\nonumber \\&+&\frac{m_j \omega_j}{ \omega_k} (\lambda_{jk}+\mu_{jk})+m_k(\lambda_{kj}-\mu_{kj})\Biggr]\coth\Biggl(\frac{\hbar\beta}{2} \sqrt{\omega_j^2-{\tilde\mu_{jj}}^2}\Biggr),
\end{eqnarray}
and
\begin{eqnarray}
 \Psi_{kj}&=&\frac{\hbar}{16}\Biggl(\frac{\omega_k \sqrt{\omega_k^2-{\tilde\mu_{kk}}^2}}{\omega_k-{\tilde\mu_{kk}}}\Biggl)\Bigl[ -\frac{1}{\omega_k}(\eta_{kj}+\nu_{kj})+ m_km_j\omega_j (\alpha_{kj} +\kappa_{kj})\nonumber \\&+&\frac{m_j \omega_j}{ \omega_k} (\lambda_{jk}-\mu_{jk})+m_k(\lambda_{kj}+\mu_{kj})\Biggr]\coth\Biggl(\frac{\hbar\beta}{2} \sqrt{\omega_k^2-{\tilde\mu_{kk}}^2}\Biggr)\nonumber \\
 &+& \frac{\hbar}{16}\Biggl(\frac{\omega_j \sqrt{\omega_j^2-{\tilde\mu_{jj}}^2}}{\omega_j-{\tilde\mu_{jj}}}\Biggl)\Bigl[ -\frac{1}{\omega_j}(\eta_{jk}+\nu_{jk})+ m_jm_k\omega_k (\alpha_{jk} +\kappa_{jk})\nonumber \\&+&\frac{m_k \omega_k}{ \omega_j} (\lambda_{kj}-\mu_{kj})+m_j(\lambda_{jk}+\mu_{jk})\Biggr]\coth\Biggl(\frac{\hbar\beta}{2} \sqrt{\omega_j^2-{\tilde\mu_{jj}}^2}\Biggr).
 \end{eqnarray}
The quantity $\Gamma_{kj}$ is given explicitly by:
\begin{eqnarray}
 \Gamma_{kj}&=&\frac{\hbar}{8}\Biggl(\frac{\omega_k-{\tilde\mu_{kk}}}{\omega_k \sqrt{\omega_k^2-{\tilde\mu_{kk}}^2}}\Biggl)\Bigl[ \eta_{kj}+\nu_{kj}- m_km_j\omega_k \omega_j (\alpha_{kj} +\kappa_{kj})\nonumber \\&+& m_k \omega_k(\lambda_{kj}+\mu_{kj})-m_j\omega_j(\lambda_{jk}-\mu_{jk})\Biggr]\coth\Biggl(\frac{\hbar\beta}{2} \sqrt{\omega_k^2-{\tilde\mu_{kk}}^2}\Biggr)\nonumber \\
 &+& \frac{\hbar}{8}\Biggl(\frac{\omega_j \sqrt{\omega_j^2-{\tilde\mu_{jj}}^2}}{\omega_j-{\tilde\mu_{jj}}}\Biggl)\Bigl[ \frac{1}{\omega_k\omega_j}(\eta_{jk}-\nu_{jk})+ m_km_j (\alpha_{jk} -\kappa_{jk})\nonumber \\&-&\frac{m_k }{ \omega_j} (\lambda_{kj}-\mu_{kj})+\frac{m_j}{\omega_k}(\lambda_{jk}+\mu_{jk})\Biggr]\coth\Biggl(\frac{\hbar\beta}{2} \sqrt{\omega_j^2-{\tilde\mu_{jj}}^2}\Biggr).
\end{eqnarray}
The diagonal elements of $\Psi$, $\Phi$ and $\Gamma$ may be derived from the above expressions by making the substitutions:
\begin{eqnarray}
 \nu_{kk}=m_k \omega_k^2, \qquad \kappa_{kk}=\frac{1}{m_k}, \qquad
 \alpha_{kk}=\eta_{kk}=0.
\end{eqnarray}
Then by solving the set of equations (\ref{set1})-(\ref{set4}), we find for the diagonal diffusion coefficients:
\begin{eqnarray}
D_{q_kq_k}&=&\frac{\hbar}{2}\Biggl(\frac{\lambda_{kk}-\mu_{kk}+\tilde\mu_{kk}}{m_k\sqrt{\omega_k^2-{\tilde\mu_{kk}}^2}}\Biggl)\coth\Biggl(\frac{\hbar\beta}{2} \sqrt{\omega_k^2-{\tilde\mu_{kk}}^2}\Biggr)\label{co1}\\
D_{p_kp_k}&=&\frac{\hbar}{2}\Biggl(\frac{m_k \omega_k^2(\lambda_{kk}+\mu_{kk}-\tilde\mu_{kk})}{\sqrt{\omega_k^2-{\tilde\mu_{kk}}^2}}\Biggl)\coth\Biggl(\frac{\hbar\beta}{2} \sqrt{\omega_k^2-{\tilde\mu_{kk}}^2}\Biggr),\label{co2}\\
D_{p_kq_k}&=&-\Biggl(\frac{\hbar \lambda_{kk}\tilde\mu_{kk}}{2\sqrt{\omega_k^2-{\tilde\mu_{kk}}^2}}\Biggl)\coth\Biggl(\frac{\hbar\beta}{2} \sqrt{\omega_k^2-{\tilde\mu_{kk}}^2}\Biggr).
\label{co3}
\end{eqnarray}
The off-diagonal elements in coordinates and momenta read
\begin{eqnarray}
D_{q_kq_j}&=&\frac{\hbar}{4}\Biggl(\frac{\lambda_{jk}-\mu_{jk}}{m_k \sqrt{\omega_k^2-{\tilde\mu_{kk}}^2}}+\frac{\tilde\mu_{kk} (\alpha_{kj}-\kappa_{kj})}{\sqrt{\omega_k^2-{\tilde\mu_{kk}}^2}}\Biggr) \coth\Biggl(\frac{\hbar\beta}{2} \sqrt{\omega_k^2-{\tilde\mu_{kk}}^2}\Biggr)\nonumber \\&+&\frac{\hbar}{4}\Biggl(\frac{\lambda_{kj}-\mu_{kj}}{m_j\sqrt{\omega_j^2-{\tilde\mu_{jj}}^2}}-\frac{\tilde\mu_{jj} (\alpha_{kj}+\kappa_{kj})}{\sqrt{\omega_j^2-{\tilde\mu_{jj}}^2}}\Biggr) \coth\Biggl(\frac{\hbar\beta}{2} \sqrt{\omega_j^2-{\tilde\mu_{jj}}^2}\Biggr),\label{co4}\\
D_{p_kp_j}&=&\frac{\hbar}{4}\Biggl(\frac{m_k\omega_k^2(\lambda_{jk}+\mu_{jk})}{ \sqrt{\omega_k^2-{\tilde\mu_{kk}}^2}}-\frac{\tilde\mu_{kk} (\eta_{kj}+\nu_{kj})}{\sqrt{\omega_k^2-{\tilde\mu_{kk}}^2}}\Biggr) \coth\Biggl(\frac{\hbar\beta}{2} \sqrt{\omega_k^2-{\tilde\mu_{kk}}^2}\Biggr)\nonumber \\&+&\frac{\hbar}{4}\Biggl(\frac{m_j \omega_j^2(\lambda_{kj}+\mu_{kj})}{\sqrt{\omega_j^2-{\tilde\mu_{jj}}^2}}+\frac{\tilde\mu_{jj} (\eta_{kj}-\nu_{kj})}{\sqrt{\omega_j^2-{\tilde\mu_{jj}}^2}}\Biggr) \coth\Biggl(\frac{\hbar\beta}{2} \sqrt{\omega_j^2-{\tilde\mu_{jj}}^2}\Biggr).\label{co5}\\
\end{eqnarray}
Finally, the elements in mixed coordinates and momenta take the form:
\begin{eqnarray}
D_{q_kp_j}&=&\frac{\hbar}{4}\Biggl(\frac{ \eta_{kj}+\nu_{kj}}{ m_k \sqrt{\omega_k^2-{\tilde\mu_{kk}}^2}}-\frac{\tilde\mu_{kk}(\lambda_{kj}+\mu_{kj})}{\sqrt{\omega_k^2-{\tilde\mu_{kk}}^2}}\Biggr) \coth\Biggl(\frac{\hbar\beta}{2} \sqrt{\omega_k^2-{\tilde\mu_{kk}}^2}\Biggr)\nonumber \\&+&\frac{\hbar}{4}\Biggl(\frac{m_j \omega_j^2 (\alpha_{kj}-\kappa_{kj})}{\sqrt{\omega_j^2-{\tilde\mu_{jj}}^2}}-\frac{\tilde\mu_{jj} (\lambda_{kj}-\mu_{kj})}{\sqrt{\omega_j^2-{\tilde\mu_{jj}}^2}}\Biggr) \coth\Biggl(\frac{\hbar\beta}{2} \sqrt{\omega_j^2-{\tilde\mu_{jj}}^2}\Biggr).\label{co6}
\end{eqnarray}
The latter formulas give the explicit analytical expressions of the diffusion coefficients, ensuring that the state of the system of harmonic oscillators evolves to the Gibbs state (\ref{gibbs}).  It can be seen that the effect of the self momentum-position coupling in the steady state rests in the renormalization of both the frequencies and the phenomenological  friction coefficients entering the expressions of the diagonal diffusion coefficients in position and momentum. For the other off-diagonal elements, the coupling constants $\tilde \mu_{kk}$ contribute in a multiplicative fashion along with the corresponding friction coefficients. 

It is worthwhile noting that   the variances with respect to the equilibrium state $\hat\rho_{\rm eq}$ are given by:
\begin{eqnarray}
 \sigma^{\rm eq}_{q_kq_k}&=&\frac{\hbar}{2 m_k\sqrt{\omega_k^2-{\tilde\mu_{kk}}^2}}\coth\Biggl(\frac{\hbar\beta}{2} \sqrt{\omega_k^2-{\tilde\mu_{kk}}^2}\Biggr),\\
 \sigma^{\rm eq}_{p_kp_k}&=&\frac{\hbar}{2}\Biggl(\frac{m_k \omega_k^2}{\sqrt{\omega_k^2-{\tilde\mu_{kk}}^2}}\Biggl)\coth\Biggl(\frac{\hbar\beta}{2} \sqrt{\omega_k^2-{\tilde\mu_{kk}}^2}\Biggr),\\
\sigma^{\rm eq}_{p_kq_k}&=&-\Biggl(\frac{\hbar \tilde\mu_{kk}}{2\sqrt{\omega_k^2-{\tilde\mu_{kk}}^2}}\Biggl)\coth\Biggl(\frac{\hbar\beta}{2} \sqrt{\omega_k^2-{\tilde\mu_{kk}}^2}\Biggr),
\end{eqnarray}
with all the other variances vanishing. This enables us to  write the diffusion coefficients in terms of the elements of the covariance matrix $\sigma^{\rm eq}$. For instance:
\begin{eqnarray}
 D_{q_kq_k}&=&(\lambda_{kk}-\mu_{kk}+\tilde\mu_{kk}) \sigma^{\rm eq}_{q_kq_k},\\
 D_{p_kp_k}&=&(\lambda_{kk}+\mu_{kk}-\tilde\mu_{kk}) \sigma^{\rm eq}_{p_kp_k},\\
 D_{q_kp_k}&=&\lambda_{kk} \sigma^{\rm eq}_{q_kp_k}.
\end{eqnarray}
Similar expressions are obtained for the remaining off-diagonal diffusion coefficients. It turns out that the latter expressions correspond exactly to the solutions of the matrix equation (\ref{maeq}), implying that $\sigma^{\rm eq}=\sigma(\infty)$, which is expected since the dynamics is assumed relaxing to $\hat\rho_{\rm eq}$, as stated above.
\subsection{Validity of the Einstein relation and quantum mechanical constraints}
By inspecting  equation (\ref{co2}) we see that when the condition
\begin{equation}
 \frac{\hbar}{2k_B T} \sqrt{\omega_k^2-{\tilde\mu_{kk}}^2}\ll 1
\end{equation}
is satisfied, the latter equation reduces to:
\begin{equation}
 D_{p_kp_k}=\Biggl(\frac{ \omega_k^2(\lambda_{kk}+\mu_{kk}-\tilde\mu_{kk})}{\omega_k^2-{\tilde\mu_{kk}}^2}\Biggl)m_k k_B T
\end{equation}
which is Einstein's relation with a frequency-dependent effective friction coefficient. The above condition is met either when the temperature is high enough, so that the behavior of the heat reservoir can be considered to be quasi-classical; it may  also be verified when 
\begin{equation}
 \omega_{kk}-\tilde\mu_{kk} \ll \frac{k_B T}{\hbar},
\end{equation}
that is when the coupling constant $\tilde\mu_{kk}$ is of the same order of magnitude as the corresponding frequency $\omega_k$.  This means that in such  physically realizable  instances, even though the temperature is low, and the quantum nature of the dynamics is dominant,   the Einstein relation, which is usually associated with the classical limit,  may  
still be valid. In that case,  the effective friction coefficient becomes very large. Since the latter is essentially positive, we infer that the following constraint holds:
\begin{equation}
 \lambda_{kk}>\tilde\mu_{kk}-\mu_{kk}.
\end{equation}

The transport coefficients displayed in equations~(\ref{coef1})-(\ref{coef2}) satisfy the  quantum-mechanical inequalities:
  \begin{eqnarray}
  &&D_{q_kq_k}D_{p_jp_j}-D_{q_kp_j}^2\ge\frac{\hbar^2}{4}\lambda_{kj}^2,\label{constr1}\\
 && D_{q_kq_k} D_{q_jq_j}-D^2_{q_kq_j}\ge\frac{\hbar^2}{4} \alpha_{kj}^2,\label{constr2}\\
 && D_{p_kp_k}D_{p_jp_j}-D_{p_kp_j}^2\ge\frac{\hbar^2}{4} \eta_{kj}^2\label{constr3},
  \end{eqnarray}
 which, mathematically speaking, result from the Cauchy-Schwartz inequality. They are direct manifestation of the fact that the Lindblad master equation preserves the fundamental properties of the density matrix. 

 From condition (\ref{constr1}), it follows that
 \begin{equation}
  \lambda_{kk}\ge\frac{|\mu_{kk}-\tilde\mu_{kk}|\omega_k}{\sqrt{\omega_k^2-{\tilde\mu_{kk}}^2}}  \cosh\Biggl(\frac{\hbar\beta}{2} \sqrt{\omega_k^2-{\tilde\mu_{kk}}^2}\Biggr).
 \end{equation}
The latter condition is always satisfied when $\mu_{kk}=\tilde\mu_{kk}$. Otherwise, in order to ensure the validity of that constraint, the temperature should verify 
\begin{equation}
 T\ge \frac{\hbar  \sqrt{\omega_k^2-{\tilde\mu_{kk}}^2}}{2 k_B  {\rm acosh}\Biggl( \frac{\lambda_{kk} \sqrt{\omega_k^2-{\tilde\mu_{kk}}^2}}{\omega_k|\mu_{kk}-\tilde\mu_{kk}|}\Biggr)},
\end{equation}
provided that
 \begin{equation}
 \lambda_{kk}\ge \frac{|\mu_{kk}-\tilde\mu_{kk}|\omega_k}{\sqrt{\omega_k^2-{\tilde\mu_{kk}}^2}}.
 \end{equation}

\section{Entanglement evolution in a Bogoliubov  bosonic system  \label{sec3}} 
In this section we apply the results obtained thus far to the Bogoliubov  Hamiltonian. The latter, which describes a system of coupled bosonic modes,   has been employed primarily to study Bose-Einstein condensate \cite{bogo}. Explicitly it reads
\begin{equation}
 \hat H=\sum\limits_{\ell, m} K_{\ell m} \hat a_\ell^\dag \hat a_m+\frac{1}{2}\sum\limits_{\ell, m} (\Delta_{\ell m}  \hat a_\ell^\dag \hat a_m^\dag+\Delta_{\ell m}^*  \hat a_\ell \hat a_m)\label{degen}
\end{equation}
where the matrix $K$ is hermetian while the matrix $\Delta$ is symmetric, i.e $K=K^\dag$, $\Delta^T=\Delta$. The bosonic operators satisfy the canonical commutation relations $[\hat a_\ell, \hat a_m^\dag]=\delta_{\ell m}$ and $ [\hat a_\ell^\dag, \hat a_m^\dag]= [\hat a_\ell, \hat a_m]=0$.  We map these operators to the canonical operators $\hat q_\ell$ and $\hat p_\ell $ through the expressions:
\begin{eqnarray}
 \hat a_\ell&=&\sqrt{\frac{K_{\ell\ell}}{2}}\Bigl(\hat q_\ell+\frac{i }{ \hbar K_{\ell\ell}}\hat p_\ell\Bigr),\\
 \hat a_\ell^\dag&=&\sqrt{\frac{K_{\ell\ell}}{2}}\Bigl(\hat q_\ell-\frac{i }{\hbar K_{\ell\ell}}\hat p_\ell\Bigr).
\end{eqnarray}
Inserting the latter expressions into the formula of the Hamiltonian (\ref{degen}) we obtain a new Hamiltonian having the same form as (\ref{orig}), with the identifications
\begin{eqnarray}
 \omega_\ell&=&\sqrt{K_{\ell\ell}^2-({\rm Re }\Delta_{\ell\ell})^2}/\hbar, \qquad m_\ell= \frac{K_{\ell\ell}}{K_{\ell\ell}-{\rm Re }\Delta_{\ell\ell}},\\
 \mu_{\ell m}&=&\frac{{\rm Im} ( \Delta_{\ell m}-K_{\ell m})}{\hbar}\sqrt{\frac{K_{\ell\ell}}{K_{mm}}}, \qquad  \nu_{\ell m}={\rm Re} (K_{\ell m}+\Delta_{\ell m})\sqrt{K_{\ell\ell}K_{mm}},\\
  \kappa_{\ell m}&=&\frac{{\rm Re} (K_{\ell m}-\Delta_{\ell m})}{\hbar^2 \sqrt{K_{\ell\ell}K_{mm}}}.
  \end{eqnarray}
  
  Notice that the system  under consideration may be of great significance in many  fields, especially in the domain of quantum information based on continuous quantum variables~\cite{brun}. Recently,  two-mode-squeezed Bose-Einstein condensate has been  employed to analyze Bell correlations and entanglement in \cite{meng}. As an other example, we may cite \cite{ritt}, which has dealt with  squeezing and entanglement in a Bose-Einstein condensate.
  
  Our goal in this work is to investigate the effect of the surrounding environment  (thermal bath)  
 on the evolution of the entanglement of  two modes ($N$=2) described by Hamiltonian~(\ref{degen}), whose initial state is a squeezed state,  the covariance matrix of which is given by
\begin{equation}
 \sigma(0)=\begin{pmatrix}
            \mathcal A (0) & \mathcal C (0)\\
            \mathcal C(0)^T & \mathcal B (0)
           \end{pmatrix} \label{initial}
\end{equation}
where
\begin{equation}
  \mathcal A (0)=\begin{pmatrix}
           \sigma_{q_1q_1}(0) & \sigma_{q_1p_1}(0)\\
           \sigma_{p_1q_1}(0) & \sigma_{p_1p_1}(0)
           \end{pmatrix}=\begin{pmatrix}
           \xi_1 & 0\\
           0 & \xi_1
           \end{pmatrix},
\end{equation}
and
\begin{equation}
  \mathcal B(0)=\begin{pmatrix}
           \sigma_{q_2q_2}(0) & \sigma_{q_2p_2}(0)\\
           \sigma_{p_2q_2}(0) & \sigma_{p_2p_2}(0)
           \end{pmatrix}=\begin{pmatrix}
           \xi_2 & 0\\
           0 & \xi_2
           \end{pmatrix},
\end{equation}
whereas
\begin{equation}
  \mathcal C (0)=\begin{pmatrix}
           \sigma_{q_1q_2}(0) & \sigma_{q_1p_2}(0)\\
           \sigma_{p_1q_2}(0) & \sigma_{p_1p_2}(0)
           \end{pmatrix}= =\begin{pmatrix}
           \theta & 0\\
           0 & -\theta
           \end{pmatrix}.
\end{equation}
The  parameters $\xi_k$ and $\theta$ are given by
\begin{eqnarray}
 \xi_1&=&n_1\cosh^2(r)+n_2\sinh^2(r)+\frac{1}{2}\cosh(2r),\\
 \xi_2&=&n_2\cosh^2(r)+n_1\sinh^2(r)+\frac{1}{2}\cosh(2r),\\
 \theta&=&\frac{1}{2}(n_1+n_2+1)\sinh(2r),
\end{eqnarray}
with $r$ being the squeezing parameter, and  $n_k$ denotes the mean thermal number of particles  in mode  $k$, namely, $n_k=\langle a_k^\dag a_k \rangle$.

In this case, the matrix $M$ takes the form  
\begin{eqnarray}
 M &&= \nonumber \\ && {\footnotesize \begin{pmatrix}
        -\lambda_{11}+\frac{{\rm Im} \Delta_{11}}{\hbar}& \frac{K_{11}-{\rm Re }\Delta_{11}} {K_{11}} & -\lambda_{12}+\frac{{\rm Im} ( \Delta_{12}-K_{12})}{\hbar}\sqrt{\frac{K_{11}}{K_{22}}}&-\alpha_{12}+\frac{{\rm Re} (K_{12}-\Delta_{12})}{\sqrt{K_{11}K_{22}}}  \\
        -K_{11}(K_{11}+{\rm Re }\Delta_{11})/\hbar^2 &-\lambda_{11}-\frac{{\rm Im} \Delta_{11}}{\hbar}&\eta_{12}-\frac{{\rm Re} (K_{12}+\Delta_{12})}{\hbar^2}\sqrt{K_{11}K_{22}} & -\lambda_{21}-\frac{{\rm Im} ( \Delta_{21}-K_{21})}{\hbar}\sqrt{\frac{K_{22}}{K_{11}}} \\
         -\lambda_{21}+\frac{{\rm Im}( \Delta_{21}-K_{21})}{\hbar}\sqrt{\frac{K_{22}}{K_{11}}}&\alpha_{12}+\frac{{\rm Re} (K_{12}-\Delta_{12})}{\sqrt{K_{11}K_{22}}}& -\lambda_{22}+\frac{{\rm Im} \Delta_{22}}{\hbar}&  \frac{K_{22}-{\rm Re }\Delta_{22}} {K_{22}} \\
       -\eta_{12}-\frac{{\rm Re} (K_{12}+\Delta_{12})}{\hbar^2}\sqrt{K_{11}K_{22}}& -\lambda_{12}-\frac{{\rm Im} ( \Delta_{12}-K_{12})}{\hbar}\sqrt{\frac{K_{11}}{K_{22}}}& -K_{22}(K_{22}+{\rm Re }\Delta_{22})/\hbar^2 &-\lambda_{22}-\frac{{\rm Im} \Delta_{22}}{\hbar}
        \label{matr2}
        \end{pmatrix}}.
\end{eqnarray}
The state of the system remains Gaussian at later times, and its covariance matrix retains its form, namely
  \begin{equation}
 \sigma(t)=\begin{pmatrix}
            \mathcal A (t) & \mathcal C (t)\\
            \mathcal C(t)^T & \mathcal B (t)
           \end{pmatrix}.
\end{equation}
\begin{figure}[htb]
{\centering{
\resizebox*{0.7\textwidth}{!}{\includegraphics{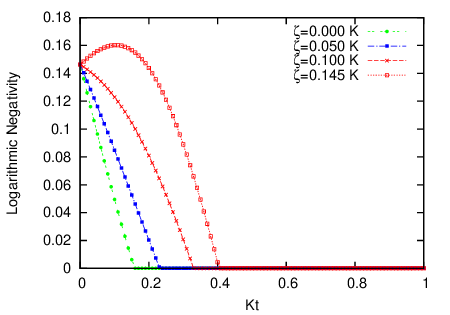}}
\par}}

\caption{\label{fig1} Evolution of the logarithmic negativity (for convenience, only the positive part is shown) as a function of time for various values of the parameter $\zeta={\rm Im} \tilde\Delta_{11}={\rm Im} \tilde\Delta_{22}$ for an initially entangled squeezed state of two identical bosons with $n_1=n_2=1$, and squeezing parameter $r=0.6$. The other parameters are: $K_{12}=K_{21}=0.2 K$, ${\rm Im }\Delta_{11}={\rm Im }\Delta_{22}=0.05 K$, ${\rm Re }\Delta_{11}={\rm Re }\Delta_{22}=0$,  ${\rm Re}\Delta_{12}=0.05 K$, ${\rm Im }\Delta_{12}=0$, $\lambda_{11}=\lambda_{22}=0.15 K$, $T=0.5 K$. All remaining parameters are set to zero, and we take $\hbar=k_B=1$ and $K=K_{11}=K_{22}$. }
\end{figure}
  This makes it possible to use the separability criteria for Gaussian states \cite{simon,adesso}
  in order to study the entanglement evolution of the system. We use the so called logarithmic negativity as a measure of separability, which is defined by \cite{isar10}
  \begin{eqnarray}
   E(\sigma(t))=-\log_2\Biggl(\det\mathcal A(t)+\det\mathcal B(t) -2 \det\mathcal C(t)\nonumber \\-
   \sqrt{\bigl(\det\mathcal A(t)+\det\mathcal B(t) -2 \det\mathcal C(t)\bigr)^2-4 \det \sigma(t)} \Biggr)
  \end{eqnarray}
  The initial state (\ref{initial}) is entangled whenever the squeezing parameter $r$ exceeds the critical value $r_c$ fulfilling $\cosh r_c=\dfrac{(n_1+1)(n_2+1)}{n_1+n_2+1}$\cite{mari}. We also suppose that the steady state is characterized by the new coupling constants $\tilde\Delta_{\ell\ell}$.
  \begin{figure}[htb]
{\centering{
\resizebox*{0.7\textwidth}{!}{\includegraphics{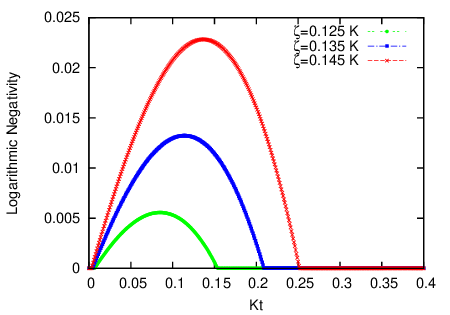}}
\par}}

\caption{\label{fig2} Evolution of the logarithmic negativity (for convenience, only the positive part is shown) as a function of time for various values of the parameter $\zeta={\rm Im} \tilde\Delta_{11}={\rm Im} \tilde\Delta_{22}$ for an initially separable squeezed state of two identical bosons with $n_1=n_2=1$, and squeezing parameter $r=0.549$.The other parameters are: $K_{12}=K_{21}=0.2 K$, ${\rm Im }\Delta_{11}={\rm Im }\Delta_{22}=0.05 K$, ${\rm Re }\Delta_{11}={\rm Re }\Delta_{22}=0$,  ${\rm Re}\Delta_{12}=0.05 K$, ${\rm Im }\Delta_{12}=0$, $\lambda_{11}=\lambda_{22}=0.15 K$, $T=0.5 K$. All remaining parameters are set to zero, and we take $\hbar=k_B=1$ and $K=K_{11}=K_{22}$.}
\end{figure}

In figure \ref{fig1}, we display the evolution of the logarithmic negativity in the course of the time for an initially entangled state for some particular values of the model. It can be seen that the fastest rate with which it suddenly vanishes occurs when the interaction coupling strengths $\tilde \Delta_{11}$ and $\tilde \Delta_{22}$ vanish in the steady state. The stronger these coupling constants are, the slower the sudden death of entanglement. We also notice that for small values, the profile is nearly linear, but  for larger values of these constants, the logarithmic negativity may even exceed its initial value, and the shape of the curve representing it deviates considerably from the linear form. 

Figure \ref{fig2} represents the logarithmic negativity for an initially separable state. It turns out that for a squeezing parameter $r$  smaller, but sufficiently close to the threshold value $r_c$, the state becomes entangled, the amount of which grows steadily with the increase of the coupling strengths in the equilibrium state. However, entanglement sudden death  occurs in all cases, but with a slower rate for strong coupling.
\section{Conclusion \label{sec4}}
We have used the Lindblad master equation, under the assumption of linear dissipation,  to  derive the explicit analytical expressions of the diffusion coefficients leading a system of coupled harmonic oscillators, weakly coupled to a heat bath, to a steady state that retains the position-momentum correlations of each oscillator. It turns out that the main effect of the latter consists in renormalizing the frequencies and the friction coefficients of the subsystems. When the physical constraints are fulfilled, we find that the validity of the Einstein relation may extend to low temperatures. We investigated  the evolution of the entanglement in a bipartite  Bogoliubov bosonic system initially prepared in a thermal squeezed state, where we find that the stronger the coupling constants are the slower the decay and sudden death occur. Entanglement generation is shown to take place for  squeezing parameter lower but sufficiently close to the critical value. These results reveal clearly that the intrinsic correlations of each subsystem, that persist in the equilibrium state, affect considerably the evolution of the other subsystems in the course of the evolution of the total system. An interesting extension of the model may consists in investigating the effect of inter-subsystem correlations and whether they have positive or detrimental effect on e.g. entanglement evolution.     

\end{document}